\documentclass{article}
\usepackage[T1]{fontenc}
\usepackage[utf8]{inputenc}
\usepackage{amsmath}
\usepackage{amssymb}
\usepackage{graphicx}
\usepackage{geometry}
\usepackage[normalem]{ulem}
\makeatletter
\usepackage{caption}
\RequirePackage{xcolor}
\definecolor{darkgreen}{rgb}{0, 0.4, 0} 
\definecolor{midgreen}{rgb}{0.5, 0.8, 0.5}
\definecolor{darkred}{rgb}{0.6, 0, 0}
\definecolor{darkblue}{rgb}{0, 0, 0.6} 

\RequirePackage{hyperref}
\hypersetup{linktocpage=true, colorlinks=true, citecolor={darkgreen}, linkcolor={darkred}, urlcolor={darkblue} }

\newcommand{\maxtime}{t_\mathrm{max}}
\newcommand{\irrev}{\,\Sigma}

\makeatother

\begin{document}

\title{{\huge\vspace{-1cm}In-vivo entropy production of \emph{A.~subaru} }}

\date{\normalsize 
1 April 2026}

\author{Yu Fu, Emmy Dobson, Benjamin B. Machta, and Michael C. Abbott
 \\ \footnotesize\itshape Physics Department \& Quantitative Biology
Institute, Yale University, New Haven, Connecticut}
\maketitle
\begin{abstract}
\noindent Entropy production is often used as a proxy for energy
consumption of a non-equilibrium system. Lower bounds can be estimated
from coarse-grained observations, and this has been done for various
biological systems. Here, we apply these tools to a more macroscopic
system whose true energy consumption is also known. We find that while
entropy production does give a lower bound, it is some 25 orders of
magnitude away from being saturated. To be certain of this result,
we survey different methods of estimating irreversibility, and write
down a novel kNN estimator.
\end{abstract}

\section{Introduction}

Life is a struggle against the inevitable increase of entropy, enabled by the consumption of free energy from the sun, or more proximally by food, or by ATP arriving from elsewhere in the cell. 
A prominent signal of such energy consumption is irreversibility: a living process looks quite different if we play the movie backwards, unlike any system at equilibrium, for which we cannot tell.
Irreversibility $\irrev$ is a statistical concept which we take to have units of bits per second.
When calculated from coarse-grained observables, it provides a lower
bound on the steady-state rate at which free energy must be consumed to drive the
system, $W$ in Joules per second \cite{esposito2012stochastic}:
\begin{equation}
k_{B}T\irrev\leq W \label{eq:irrev-work}
\end{equation}
where  $k_{B}=1.38\times10^{-23}\,$J/K, and $T$ is the temperature.

Yet not everything in life is driven. For instance our hair contains
no muscles, although our arms do. One application of these ideas to
biology is as a way to demonstrate that certain microscopic protrusions
from a bacterium are more like arms than hairs. An observation of
nonzero $\irrev$ proves that the system must be powered \cite{battle2016broken}, and such
a proof holds even if we observe only a few of the degrees of freedom
\cite{roldan2010estimating}.
For more macroscopic living systems,
we are seldom in doubt about whether they are driven, but we
can still estimate their irreversibility from data. What can be learned
in such cases?

\begin{figure}
    \centering \includegraphics[width=14cm]{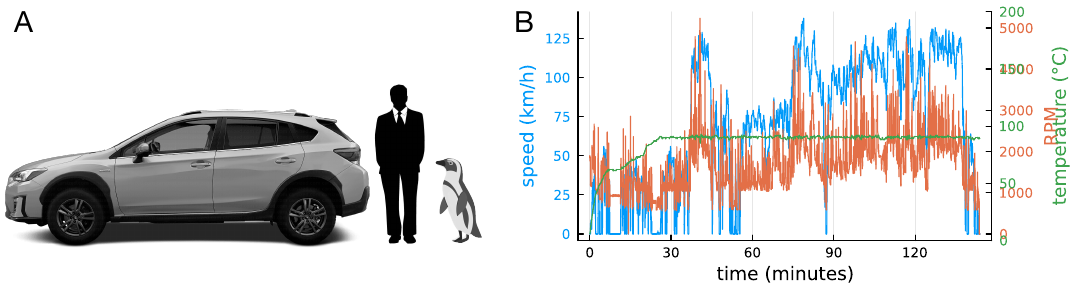}\caption{(A) \emph{Automobilus subaru} (left) is a large quadrirotus creature, originally
    from Japan. Shown with \emph{H.~sapiens} (centre) and \emph{S.~demersus}
    (right) for scale. (B) Raw data time-trace, showing speed, engine
    RPM, and coolant temperature. These three numbers are recorded about every 140\,ms, using a device \cite{huang00freematics} which plugs into the OBD-II port \cite{epa1993obdii}.
    We regard the temperature as a constant, and use the other two as $x(t) \in \mathbb{R}^2$ for our EPR calculations (figures \ref{fig:k-means}, \ref{fig:many-methods}).
    }\label{fig:the-system}\label{fig:raw-data}
    \end{figure}

Inspired by Bennett~et.~al.~\cite{bennett2009neural}, 
this letter explores
the meaning of macroscopic entropy production by studying experimental
data from a rather large model organism, shown in figure \ref{fig:raw-data}.
Of course we do not observe all of its degrees of freedom, but working
from what we do have, we can estimate the entropy production (using methods described in results) to be
at least
\begin{equation}
\irrev\approx0.5\,\text{bits/s}.\label{eq:onenumber}
\end{equation}
Unlike some microscopic biological systems, here we have little doubt
that this system is driven --- not only as a matter of natural law,
but also that of the state legislature \cite{connecticut2017autonomous}. We can estimate its rate
of work by measuring its diet (as described in the methods section), and converting to SI units we have:
\begin{equation}
k_{B}T\irrev\approx2\times10^{-21}\,\text{J/s}\qquad\leq W\approx7\times10^{4}\,\text{ J/s}.\label{eq:bound}
\end{equation}
We observe that the inequality is quite far from being saturated.
The gap obtained in equation \eqref{eq:bound} is, to our knowledge,
an increase on any so far reported in the literature. A gap of 8
orders of magnitude was reported by \cite{pietzonka2024thermodynamic}
 for neurons, and  4 orders of magnitude in \cite{roldan2021quantifying}, studying
auditory hair cells.
We can compute similar gaps for many other biological systems where a paper has reported something like $\irrev$, by estimating their power $W$ by independent means.
The resulting comparisons are shown in figure \ref{fig:Comparison}.

The next section looks at how we arrived at equation \eqref{eq:onenumber} from data, and the uncertainties
of its estimation. We return to questions of biology in the conclusion.

\section{Results}

Because so many ways of estimating irreversibility have been proposed,
we thought it might be a service to the field to apply many of them
to the same system. As is traditional, we also introduce our own,
superior, estimator, which we presume will soon be widely adopted. Alongside each formula, we give a numerical estimate read off figure \ref{fig:many-methods}; these involve some choices described below.

Suppose that $X(t)$ is a random variable taking values $x\in\mathbb{R}^d$ at each time $0\leq t\leq\maxtime$.
If we write its time-reverse as $\tilde{x}(t)=x(\maxtime-t)$, then
the statistical irreversibility is defined as the KL divergence between
forward and reversed trajectories:
\begin{equation}
\irrev=\lim_{\maxtime\to\infty}\frac{1}{\maxtime}D_{\mathrm{KL}}\big[p(X)\big\Vert p(\tilde{X})\big].
\label{eq:defn-irrev}
\end{equation}
This quantity is often called the entropy production rate (EPR). The true microscopic EPR will saturate \eqref{eq:bound}, and we write this as $\dot{S}$. The irreversibility $\irrev$ calculated from macroscopic variables may be viewed as one way to estimate a coarse-grained EPR. But there are other such estimates, independent of irreversibility. Below we calculate one, $\dot{S}_\mathrm{TUR}$, and discuss some others. 

The difficulty of estimating irreversibility $\irrev$ from data is that the space of possible
trajectories integrated over in \eqref{eq:defn-irrev} is extremely sparsely covered. One solution is to assume
a data-generating process for which we can calculate $p(X)$ everywhere,
and then fit the parameters of this process to the data. 

The first such method we use assumes Gaussian statistics, with continuous
space and time. For this case, Seara~et~al.~\cite{seara2021irreversibility}
derived:
\begin{equation}
\irrev_{\mathrm{gauss}}=\int\frac{d\omega}{4\pi}\mathop{\mathrm{Tr}}\left\{ \left[C^{-1}(-\omega)-C^{-1}(\omega)\right]C(\omega)\right\} \qquad\Rightarrow0.5\,\text{bits/s}.\label{eq:gaussSdot}
\end{equation}
Here $C(\omega)$ is the $d\times d$ covariance matrix, and $\omega$ is frequency.
If the covariance is estimated from a finite time-series, this formula
will tend to over-estimate the true entropy production. The method
of \cite{seara2021irreversibility} uses a Gaussian smoothing $\sigma$
in frequency space, for which they were able to calculate the $\irrev_{\mathrm{gauss}}$
for uncorrelated noise, finding $\irrev_{\mathrm{bias}}=d(d-1)N\sqrt{\pi}/(2\maxtime^{2}\sigma)$
for $N$ samples. Subtracting this bias leads to an improved estimate. 

    \begin{figure}
    \centering \includegraphics[width=14cm]{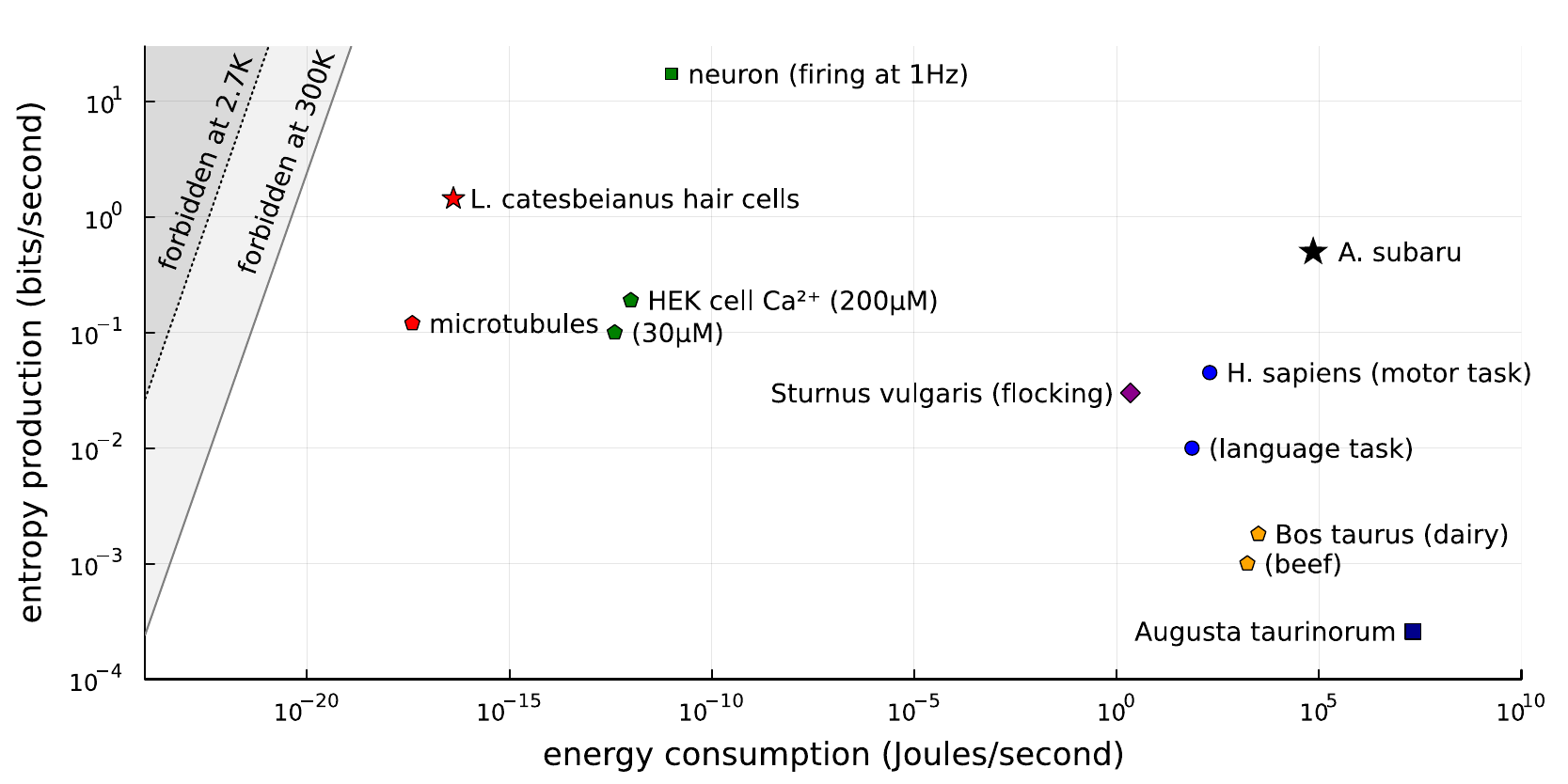}\caption{Entropy production provides a bound on the rate of energy consumption, at
    a given temperature. Besides our work (black star), we plot some competing papers; only the one on \emph{L.~catesbeianus} hair cell bundles \cite{roldan2021quantifying} provides figures for both axes (red star). For all other points, we have estimated energy consumption (as described in Methods) and plot irreversibility / entropy production rates taken from the following papers: \emph{Bos taurus}~\cite{skinner2021estimating}, \emph{H.~sapiens}~\cite{lynn2021broken}, \emph{Sturnus vulgaris}~\cite{ferretti2022signatures}, microtubules and HEK cells~\cite{skinner2021improved}, neurons~\cite{pietzonka2024thermodynamic}, 
    and the Turin public transit system~\cite{biazzo2020city}.
    The two red points are sub-cellular scale, while the green points are single cells.
    }\label{fig:Comparison}
    \end{figure}

\begin{figure} 
    \centering \includegraphics[width=10cm]{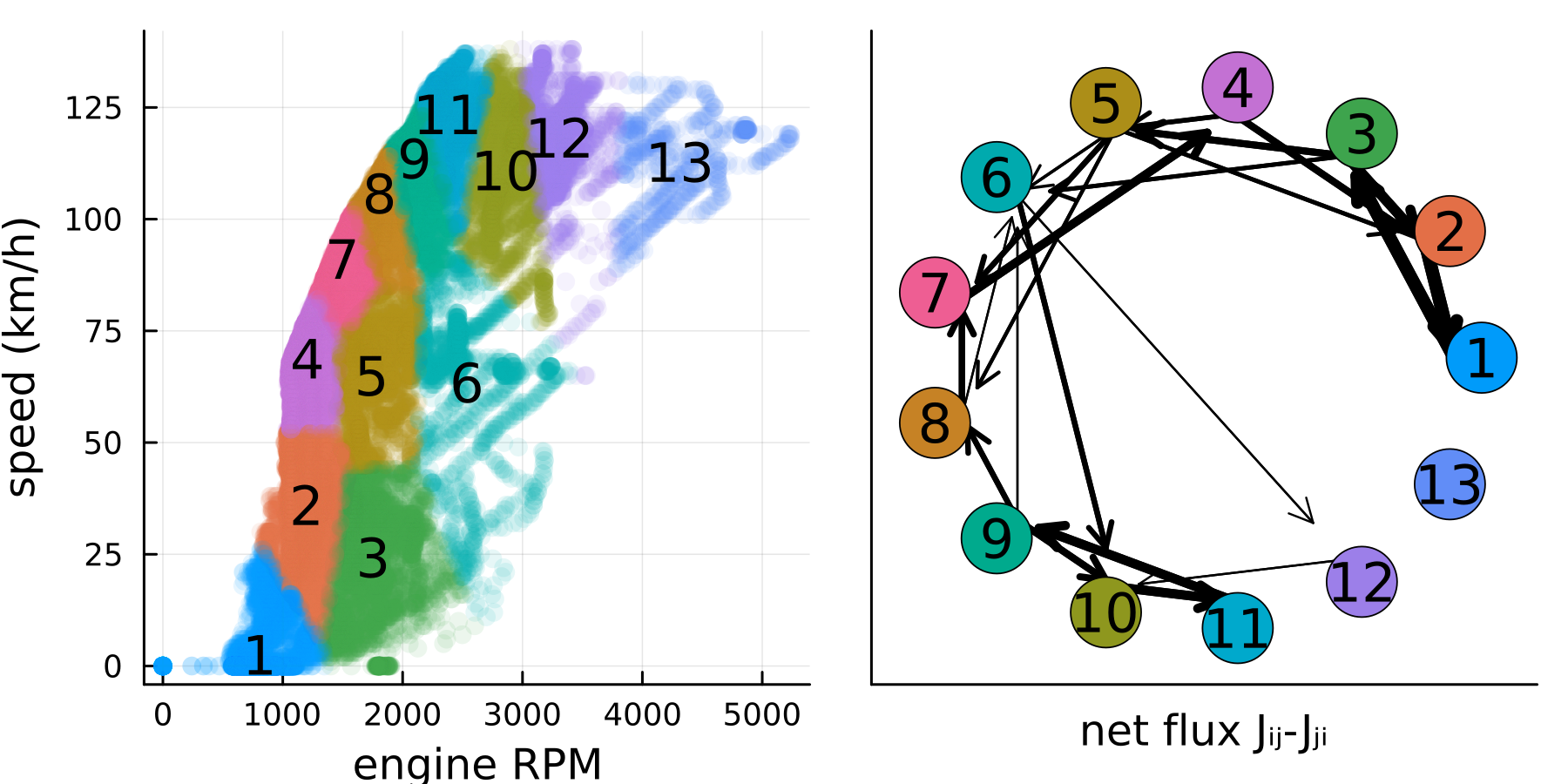}\caption{Clustering data using k-means to compute irreversibility. Same data as figure \ref{fig:raw-data},
    with $k=13$ clusters. Discretising on a grid instead would result in many
    empty squares. (We observe that this \emph{subaru} has a CVT \cite{speedkar99subaru}, as fixed gear
    ratios would show up as radial lines on this plot.) Right, width of
    arrows is $J_{ij}-J_{ji}$, showing non-equilibrium fluxes for 1st-order
    Markov formula (\ref{eq:histogramSdot}). }\label{fig:k-means}
    \end{figure}

\begin{figure}
    \centering \includegraphics[width=16cm]{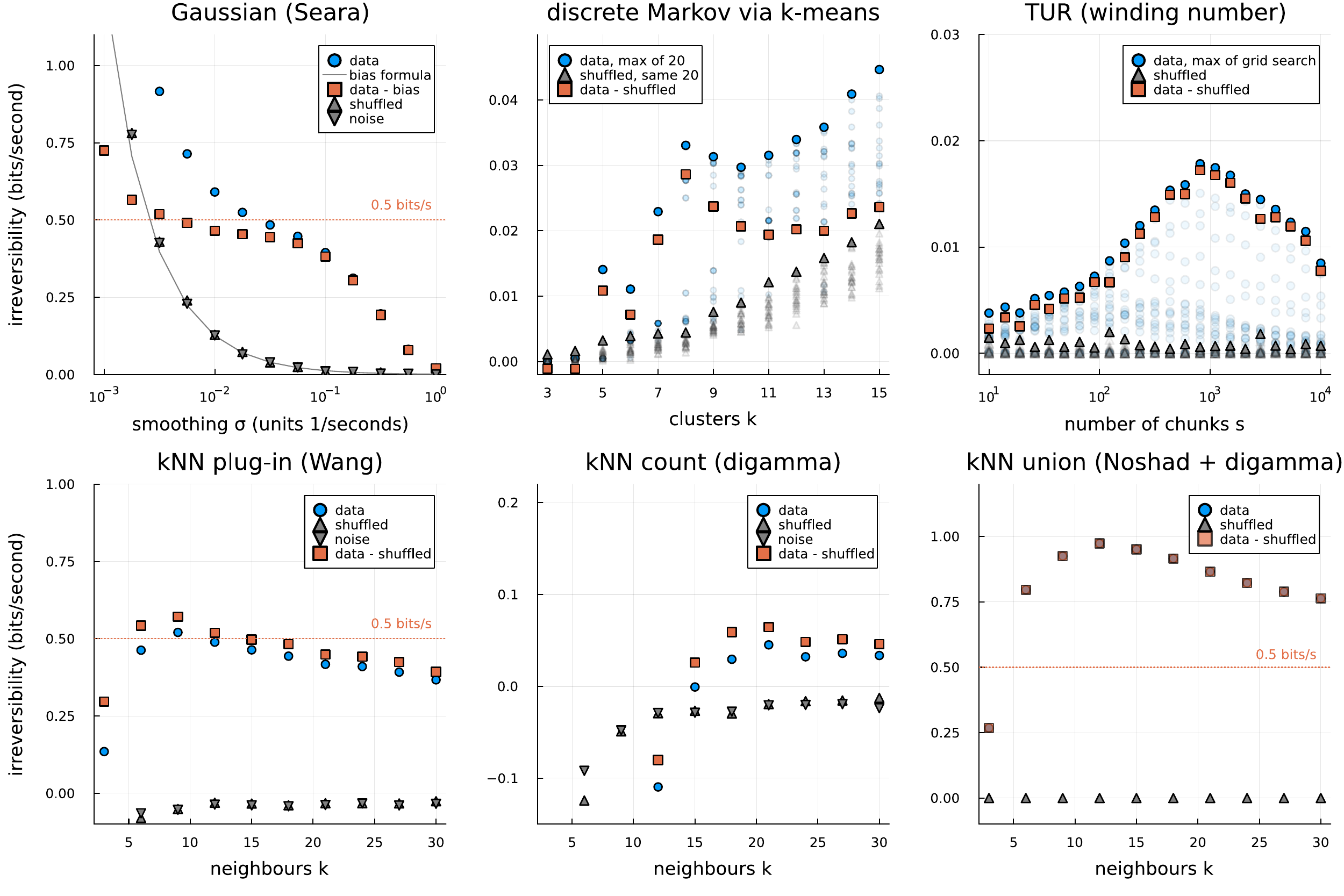}
    
    \caption{The effect of hyper-parameters and bias on several methods of estimating
    entropy production, equations (\ref{eq:gaussSdot}) to (\ref{eq:turSdot}). All estimates from data (blue points) are deterministic, except
    for (\ref{eq:histogramSdot}) where k-means clustering has a random
    start, hence we plot 20 repetitions of the clustering algorithm, and
    the maximum. The shuffling of data, and the noise, are also random (grey points). 
    On all plots, orange points are estimates corrected for bias; these are closer to being independent of the hyper-parameter. The dotted orange line at $\irrev=0.5\,$bits/s is the value in \eqref{eq:onenumber}, and used in figure \ref{fig:Comparison}.
    For the TUR estimate
    (\ref{eq:turSdot}), we compute winding about many points $x$, on
    a grid, and highlight the maximum.
    All plots use
    the same data, with $d=2$ and $N=134,600$ points over $\maxtime=324$ minutes, a longer time-series like that shown in figure \ref{fig:raw-data}B.
    }\label{fig:many-methods}
    \end{figure}

Another simple data-generating process is a Markov chain, with discrete
time-steps, and the simplest case to consider has discrete state
space too. If the flux from state $i$ to state $j$ is $J_{ji}$,
then the irreversibility is:
\begin{equation}
\irrev_{\text{discrete}}=\sum_{i<j}(J_{ij}-J_{ji})\log\frac{J_{ij}}{J_{ji}}\qquad\Rightarrow0.03\,\text{bits/s}.\label{eq:histogramSdot}
\end{equation}
Continuous data can be clustered into $k$ discrete states, and for
small enough $k$, most transitions will be observed, and thus we
obtain a simple empirical estimate of the transition rates.
This method was used by \cite{lynn2021broken}, and applied to our data, the clustering and the resulting fluxes $J_{ij}$ are shown in figure \ref{fig:k-means}.
Such methods will also tend to produce finite estimates on noise. The formula we derived for this is a poor match for the data, so in figure \ref{fig:many-methods} we compute the bias using transition rates from shuffled data, for which the true irreversibility must be zero.

\begin{figure}
    \centering \includegraphics[width=8cm]{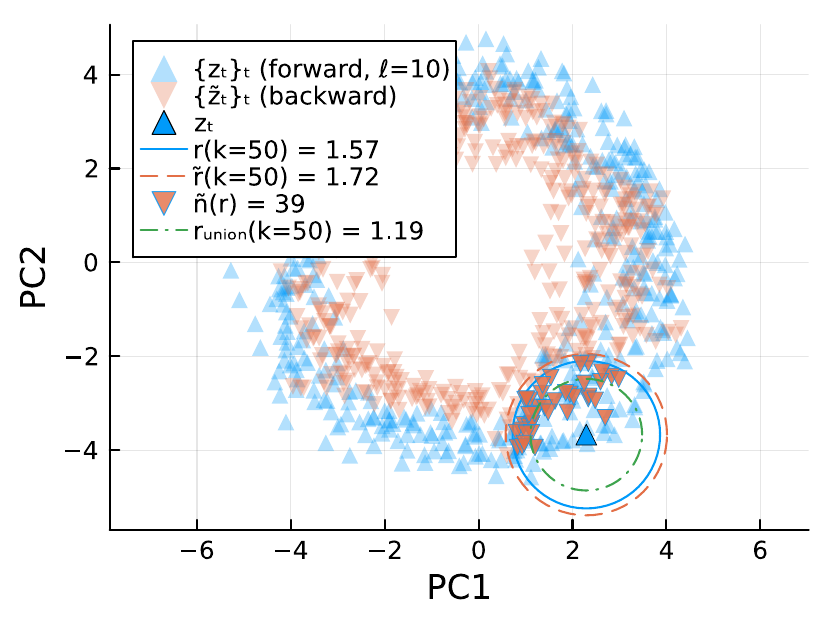}\caption{Sketch of three kNN irreversibility estimators, using a toy dataset with $N=500$ points
    in $d=2$ dimensions. We perform a 10th order Markov estimate, hence
    $z_{t}=(x_{t},x_{t+1},\ldots,x_{t+\ell})\in\mathbb{R}^{20}$; the
    plot shows two principal components. For a given point $z_{t}$, the
    radii of the circles containing $k$ forward points (blue), or $k$
    reverse points $\tilde{z}_{t'}$ (orange), are the input to the plug-in
    estimator (\ref{eq:wangSdot}). For the KSG-like estimator  $\irrev_\text{count}$ from (\ref{eq:digammaSdot}),
    instead the number of orange points within the blue circle is what
    matters. Finally, $\irrev_\text{union}$ from (\ref{eq:noshadSdot}) finds distance to the $k$-th
    nearest point among the union of all $z_{t}$ and $\tilde{z}_{t}$,
    and then counts both blue and orange points within this green circle.
    }\label{fig:kNN-sketch}
    \end{figure}

Instead of discretising space, there are various $k$ nearest neighbour
(kNN) approaches for esimating many information-theoretic quantities from data. Defining $z_{t}=(x_{t},x_{t+\delta t})\in\mathbb{R}^{2d}$
and $\tilde{z}_{t}=(x_{\maxtime-t},x_{\maxtime-t-\delta t})$ makes
it easy to use the KL divergence estimator from Wang~et.~al.~\cite{wang2009divergence} to compute irreversibility.
If $\epsilon_{k}(z)$ is the distance to the $k$-th nearest $z$
vector, and $\tilde{\epsilon}_{k}(z)$ to the $k$-th nearest $\tilde{z}$
vector, then approximating $p(z)\propto1/\epsilon_{k}(z)^{2d}$ leads
to:
\begin{equation}
\irrev_{\text{plug-in}}=\frac{1}{\maxtime}\sum_{t}2d\log\frac{\tilde{\epsilon}_{k}(z_{t})}{\epsilon_{k}(z_{t})}\qquad\Rightarrow0.5\,\text{bits/s}.\label{eq:wangSdot}
\end{equation}
This estimator was used by \cite{watson2018method_alt, tan2021scaledependent} to calculate irreversibility.
The other class of kNN estimators counts points within a ball instead,
following the work on estimating mutual information by KSG \cite{kraskov2004estimating}. We have never seen this idea
applied to irreversibility, but the formulae are simple enough to derive.
If $\tilde{n}(z,\epsilon)$ counts points for which $\left|\tilde{z}_{t}-z\right|<\epsilon$,
then our first formula is
\begin{equation}
\irrev_{\text{count}}=\frac{1}{\maxtime}\sum_{t}\psi(k)-\psi\big(\tilde{n}(z_{t},\epsilon_{k}(z_{t}))+1\big)\qquad\Rightarrow0.05\,\text{bits/s}\label{eq:digammaSdot}
\end{equation}
where $\psi$ is the digamma function. But what performs better is the following form, adapted from KL divergence estimators of Noshad~et.~al.~\cite{noshad2017direct}. 
If $\epsilon_{k}^{\cup}$ is the distance to the $k$-th nearest
neighbour among all points forward \& reverse, $\{z_{t}\}_{t}\cup\{\tilde{z}_{t}\}_{t}$, then we define:
\begin{equation}
\irrev_{\text{union}}=\frac{1}{\maxtime}\sum_{t}\psi\big(n_{t}(z_{t},\epsilon_{k}^{\cup})+1\big)-\psi\big(\tilde{n}_{t}(z_{t},\epsilon_{k}^{\cup})+1\big)\qquad\Rightarrow0.8\,\text{bits/s}.\label{eq:noshadSdot}
\end{equation}
Figure \ref{fig:kNN-sketch} below sketches these KSG-like estimators, using toy data.

Without computing irreversibility at all, there are still other independent ways to to bound the entropy production $\dot{S}$.
The measure we calculate here comes from what is known as the thermodynamic uncertainty relation (TUR), which we write as $k_B T \dot{S}_{\mathrm{TUR}} \leq W$. It  requires that we pick an observable $F$, and then calculates a bound based on the variance of that quantity \cite{barato2016cost, gingrich2016dissipation}:
\begin{equation}
\dot{S}_{\mathrm{TUR}}(F)=\frac{2}{\maxtime}\frac{\left\langle F\right\rangle ^{2}}{\mathop{\mathrm{Var}(F)}}\qquad\Rightarrow0.02\,\text{bits/s}.\label{eq:turSdot}
\end{equation}
We could take $F$ to be, for instance, the winding number about the
mean, if the data is in $d=2$ dimensions. To estimate this from one
long time-series, we divide the data into $s$ pieces, each of duration
$\maxtime/s$. 

We now turn to applying the above formulae to our experimental data on \emph{A.~subaru}. 
Since each method involves a choice of some hyper-parameter, figure
\ref{fig:many-methods} shows graphically the effect of varying the smoothing $\sigma$, the number of clusters or neighbours $k$, or the number of chunks $s$.
All use the same data ($N=134,600$ points from $\maxtime=324$ minutes,
$d=2$), and all plots show results from formulas above in blue.
They also all show noise estimates, from either shuffling the time-points or simply generating Gaussian
noise, in grey. Orange points are estimate minus
noise, and are less strongly affected by varying $\sigma,k,s$.
All estimators are clearly nonzero, but the values vary quite widely.
The orange dotted line represents our rough consensus estimate of 0.5\,bits/s, \eqref{eq:onenumber} above.

\begin{figure}
    \centering \includegraphics[width=11cm]{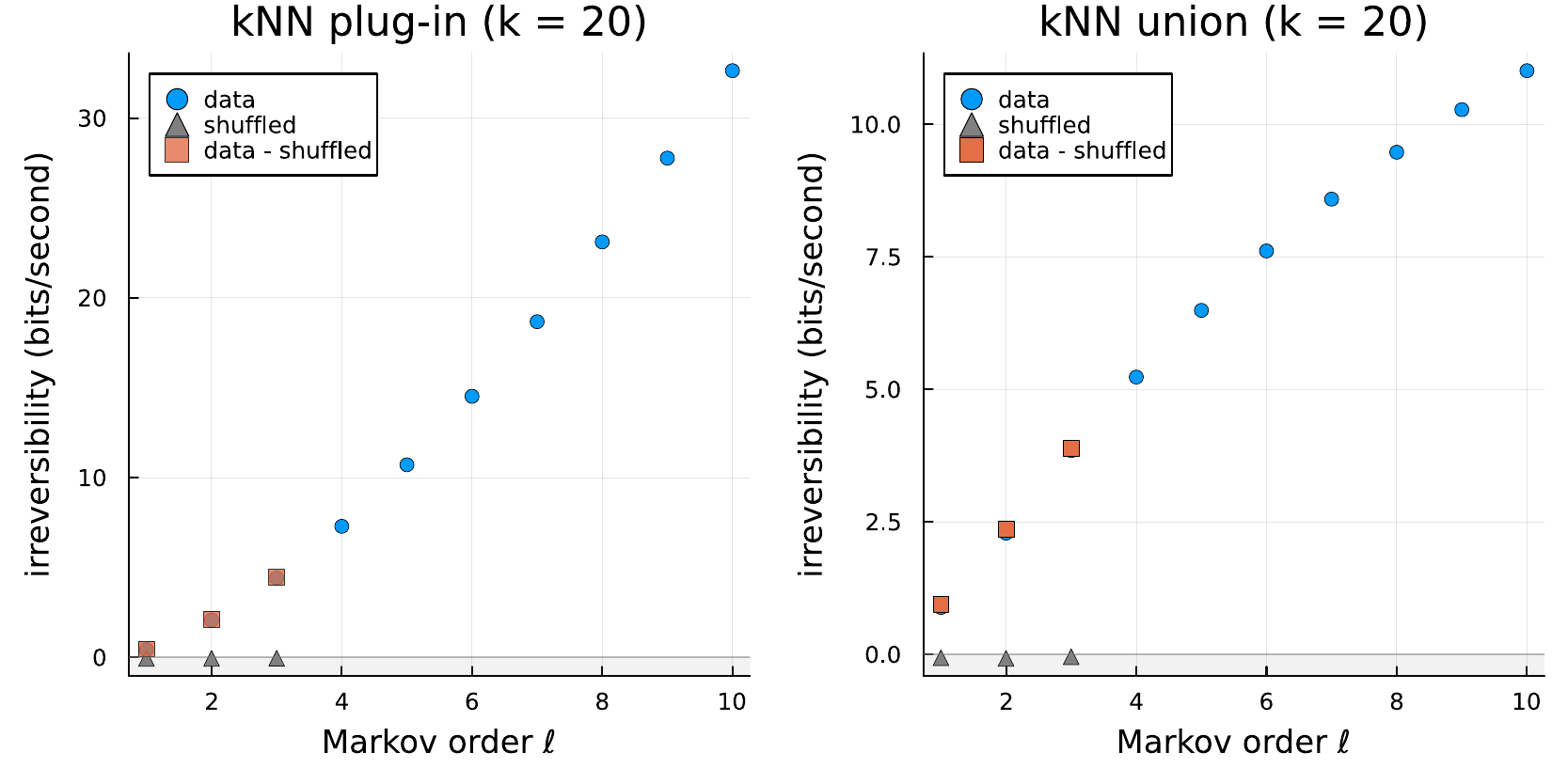}
    \caption{The effect of Markov order $\ell$ on two kNN methods of estimating irreversibility, \eqref{eq:wangSdot} and \eqref{eq:noshadSdot}.
    The increase with $\ell$ means that most of the physics is far from the sampling time-scale $\delta t \approx 140\,$ms, perhaps unsurprisingly. (The shuffled
    points stop early, as these become very slow to compute.)
    Attempts to compute similarly high-order versions of the discrete Markov estimator \eqref{eq:histogramSdot} fail, as there is insufficient data to estimate the fluxes on all possible transition paths.
    }\label{fig:markov-order}
    \end{figure}

Any of the Markov estimators can also be extended to higher-order
Markov processes, that is, to allow dependence on $\ell$ time-steps
of history. While discrete histogram method generalising \eqref{eq:histogramSdot} will suffer, as the $k^{2}$
fluxes $J_{ij}$ are replaced by $k^{\ell+1}$ fluxes $J_{ij\ldots m}$,
the kNN methods are easily adapted by writing $z_{t}=(x_{t},x_{t+\delta t},\ldots,x_{t+\ell\delta t})\in\mathbb{R}^{\ell d}$.
Figure \ref{fig:markov-order} shows what happens on our data: we
see a nearly linear increase as far as we are able to calculate.

There are other statistical EPR approaches which work quite differently, and do not involve calculating irreversibility $\irrev$ at all.
Above we already introduced the TUR.
Figure \ref{fig:Comparison} also has datapoints from Skinner \& Dunkel who define a waiting-time estimator \cite{skinner2021estimating}.
A motivating case is to consider that we observe only whether a metronome is left or right of centre.
With two discrete states clearly $\irrev_{\text{discrete}}=0$ as there can be no closed cycles.
But in fact a regular metronome cannot be made from equilibrium
components, and the estimate of \cite{skinner2021estimating} asks for the cheapest model which reproduces the observed waiting time distribution.
However, it's not clear how to produce a meaningful waiting time distribution from our dataset.

There are also ways to bound the energy consumption of a system which do not rely on statistical information alone. For instance, to apply the method of \cite{leighton2024jensen} we would need to know the friction coefficient. Such approaches are arguably half way between statistical $\irrev$ and physical $W$, and are beyond the scope of this letter.

\section{Conclusion}

How should we interpret measurements of macroscopic irreversibility, 
or coarse-grained entropy production rates?
We believe it's important to ask whether the degrees of freedom 
observed capture the biological purpose of the system. 

For neurons, it is not unreasonable to suppose that their purpose
is to produce spikes, or to convey them from one part of the body
to another. All their other complicated cellular processes are not
just unobserved, but also things that evolution would presumably be
happy to dispense with, if it could. The fact that it cannot, that
it spends 8 orders of magnitude more ATP than is apparently required
\cite{laughlin1998metabolic}, is interesting as it implies some other
constraint must be binding. (In a slightly different context, the
Landauer bound of $1\,k_{B}T$ per bit \cite{landauer1961irreversibility, bennett1982thermodynamics}
is often thought of as a fundamental limit on computing, and the gap
between this and real computers is our room to improve.)
Clocks have a similarly clear purpose, as the desired
output is surely a periodic signal. There is a thermodynamic bound \cite{barato2016cost},
and artificial clocks get close to it \cite{pearson2021measuring}.
If natural circadian clocks cannot do so, then evolution
must be pushing against some other constraint.

However, in many other systems, the degrees of freedom which are accessible
to measurement are not the ones the organism cares about. The purpose
of \emph{A.~subaru}'s drivetrain is largely to push air out of the
way, which consumes (by biological standards) vast amounts of energy.
The purpose of \emph{E.~coli}'s flagella is likewise to push its body
through water, not merely to produce regular circular motion. In either
case, a motor which produces more- or less-irreversible rotational
data isn't better or worse. Biology surely cares about MPG but not $\Sigma$ here.

The irreversibility seen in the coarse-grained variables which we
can measure in such macroscopic systems may still be a useful summary
statistic, but its usefulness has to rest on grounds other than the
relation to energy, equation \eqref{eq:irrev-work}. It may be true, for instance,
that the regularity of our heartbeat or brain waves, or the circulation of a flock
of birds or a swarm of traffic, is indicative of some interesting
feature. (Beyond, one hopes, whether the local laws of the road follow
the Buddhist sign convention or the B\"{o}n one \cite{karmay1975general}.) But any correlation
with energy consumption, 25 orders of magnitude away, is likely to
be a co-incidence. There has to be an independent reason to like this
statistic.

On a less exalted level, macroscopic data is surely no worse than
computer-generated data as a test case for understanding the biases
and limitations of statistical tools for computing irreversibility.
Comparisons between different methods appear to be rare in the literature, and the wide range of values seen in figure \ref{fig:Comparison} may be surprising.
Our appendix (which wasn't ready in time for v1) looks at a few more.

\section*{Acknowledgements}

We thank Jose Betancourt, Pranav Kantroo, Matthew Leighton, Christopher Lynn, Diana Valverde Mendez, Asheesh Momi, Mason Rouches, XJ Xu, and Qiwei Yu for helpful conversations. 
Thierry Emonet requests that we note that we have kept no record of
who was driving down I-84 on 10 April 2025 (figure \ref{fig:raw-data}B).


\section*{Methods}

Figure \ref{fig:Comparison} plots data from many sources.  
Note that many papers measure entropy production rates in $k_{B}\cdot\text{nats/s}$, which is $k_{B}\log(2)$ times what we call the EPR, in bits/s.

\paragraph{\emph{Automobilus subaru}}
All data was recorded from the OBD-II port \cite{epa1993obdii} using a Freematics device \cite{huang00freematics}. An example trace of raw data is shown in figure \ref{fig:raw-data}B.
For the energy calculation, 1 US gallon of petrol is about 130\,MJ. If the \emph{subaru} gets 30 miles per gallon at 60 miles per hour, then its power is 72\,kW thermal.

\paragraph{\emph{Bos taurus}}
We take the entropy production rate from
\cite{skinner2021estimating}, computed from a sequence of sitting and standing times of the whole animal.
Dairy cows eat about 55 lbs dry matter per day, 11\,MJ metabolisable
energy per kg, giving 3\,kW power. Cows on the beef career track eat
less, about 30\,lbs/day, giving 1.7\,kW \cite{stpierre2003economic}. 

\paragraph{\emph{Sturnus vulgaris}} The entropy production flocking we read from \cite{ferretti2022signatures} figure 1(j), giving 0.03 bits/s per bird. For energy consumption, \cite{westerterp1985energetic} gives 190 kJ/day which is $2.2\,\mathrm{W}$. 

\paragraph{\emph{Augusta~taurinorum}}
The irreversibility of Turin's public-transport flow is estimated from the asymmetry between morning and afternoon commuting fluxes, following the analysis of \cite{biazzo2020city}. The KL divergence between morning and afternoon commuting fluxes in Turin 
is $D_{\mathrm{KL}} \approx 0.64$ nats. Assuming a characteristic commuting time $T_{\text {travel }} \approx 1$ hour $=$ 3600 s , the corresponding irreversibility is estimated as $\Sigma \approx D_{\mathrm{KL}}/(2 T_\text {travel } \log 2) \approx 2.6 \times 10^{-4}$ bits s $^{-1}$. A March 2026 report on a Turin city council commission hearing \cite{Zanna2026Pubblico} stated that in 2025 GTT vehicles travelled about 55.1 million km and carried about 300 million passengers. We thus inferred an annual fuel consumption of $6.28 \times 10^{15} \mathrm{~J}$, corresponding to an average power of $2.1 \times 10^7 \mathrm{~J} / \mathrm{s} \approx 21 \,\mathrm{MW}$.

\paragraph{\emph{Homo sapiens}} The irreversibility in the brain under different physical and cognitive activities (language and motor tasks) is taken from Fig.~4A of \cite{lynn2021broken}. Humans require about $1500 \mathrm{~kcal}$ per day, corresponding to $73 \mathrm{~W}$. Energy expenditure during exercise varies widely; in the plot, we use $73 \mathrm{~W}$ for language and 200 W for motor activity. We argue that using whole-body energy consumption as an approximation for brain energy consumption gives the correct order of magnitude, within about two orders of magnitude.

\paragraph{\emph{Microtubulus}} Microtubule dynamics are powered by GTP hydrolysis, with approximately one GTP consumed per tubulin dimer added to the growing lattice. Walker et al.\cite{walker1988dynamic} report a typical plus-end growth speed of \(v \approx 25\,\mathrm{nm\,s^{-1}}\) in vitro. For a 13-protofilament microtubule, one added layer corresponds to a length increase of \(\ell = 8\,\mathrm{nm}\) across all 13 protofilaments, so the GTP consumption rate is estimated as \(\dot{n}_{\mathrm{GTP}} = (v/\ell)\times 13 = (25/8)\times 13 \approx 40\,\mathrm{s^{-1}}\) per microtubule. Taking the free energy of GTP hydrolysis under physiological conditions to be \(\Delta G \approx -54\,\mathrm{kJ\,mol^{-1}}\), or roughly \(21\,k_B T\) per event, this corresponds to an average energy consumption \(W \approx 40\,\mathrm{s^{-1}} \times 21\,k_B T \approx 4\times 10^{-18}\,\mathrm{W}\) per microtubule.The "irreversibility" is from the \(\sigma_2\) bound in \cite{skinner2021improved}: \(\irrev \ge 5\,k_B\,\mathrm{min^{-1}} \approx 0.12\,\mathrm{bits\,s^{-1}}\).

\paragraph{\emph{Lithobates catesbeianus} hair-bundle oscillations}
The irreversibility and energy consumption of hair-bundle oscillations are converted from Fig 5c,d of \cite{roldan2021quantifying} respectively.

\paragraph{\emph{Neuronum}}
The irreversibility for neuronal spike trains is taken from \cite{pietzonka2024thermodynamic}, which gives a lower bound of about $12\,k_B$ per spike for purely excitatory stimulation. The total energy required to generate an action potential is estimated to be about $10^{-11}\,\mathrm{J}$ per spike, given in the same paper. 

\paragraph{HEK cell \protect\boldmath$\mathrm{Ca}^{2+}$ oscillations} We estimate energy consumption using the fact that a typical mammalian cell dissipates about $10^{-12}-10^{-11} \mathrm{~W}$ in total. Since $\mathrm{Ca}^{2+}$ oscillations are only one part of the cell's metabolism, we assign them an energetic cost of
$W_{\mathrm{Ca}} \sim 10^{-13}-10^{-12} \mathrm{~W}
$ per cell. We use $10^{-12}\mathrm{~W}$ for HEK cells at $30\,\mu$M carbachol and $4 \times 10^{-13}\mathrm{~W}$ at $30\,\mu$M carbachol. The irreversibility is from the \(\sigma_2\) bound in \cite{skinner2021improved}.

\bibliographystyle{my-JHEP-5.bst}
\bibliography{My-Library, extra, fig2}

\end{document}